\documentclass{PoS}

\PoS{PoS(LAT2005)048}

\newcommand{\be}{\begin{equation}}
\newcommand{\ee}{\end{equation}}

\newcommand{\chiPT}{$\chi$PT}
\newcommand{\FRRchiPT}{FRR $\chi$EFT}
\newcommand{\FRRchiEFT}{FRR $\chi$EFT}
\newcommand{\chiEFT}{$\chi$EFT}

\newcommand{\mpi}{m_\pi}
\newcommand{\fpi}{f_\pi}

\newcommand{\gev}{\,{\rm GeV}}

\title{Power Counting Regime of Chiral Extrapolation\\ and Beyond }

\ShortTitle{Power Counting Regime of Chiral Extrapolation and Beyond }

\author{\speaker{Derek B. Leinweber}
        \thanks{This work is supported by the Australian Research
                Council and by DOE contract DE-AC05-84ER40150, under
                which SURA operates Jefferson Laboratory.}
        \\
        Special Research Center for the Subatomic Structure of Matter,
        and
        Department of Physics, University of Adelaide, 
        Adelaide SA  5005  Australia\\ 
        E-mail: \email{dleinweb@physics.adelaide.edu.au}}

\author{Anthony W. Thomas and Ross D. Young\\
        Jefferson Lab, 12000 Jefferson Ave., Newport News, VA 23606,
        USA \\
        E-mail: \email{awthomas@jlab.org and young@jlab.org}}

\abstract{Finite-range regularised chiral effective field theory is
  presented in the context of approximation schemes ubiquitous in
  modern lattice QCD calculations.  
  Using FRR techniques, the power-counting regime
  (PCR) of chiral perturbation theory can be estimated.  To
  fourth-order in the expansion at the 1\% tolerance level, we find $0
  \le m_\pi \le 180$ MeV for the PCR, extending only a small distance
  beyond the physical pion mass.  
}

\FullConference{XXIIIrd International Symposium on Lattice Field Theory\\
		 25-30 July 2005\\
		 Trinity College, Dublin, Ireland}

\begin{document}

\section{Introduction}

Most everyone is now familiar with the importance of accounting for
chiral nonanalytic behavior in the quark-mass extrapolation of
physical observables.  The (pseudo) Goldstone bosons associated with
dynamical chiral symmetry breaking in QCD couple strongly and give
rise to a quark-mass dependence of hadron observables in which
significant curvature is usually encountered in approaching the
physical regime.  The Adelaide Group has played a leading role in
emphasizing the role of this physics
% in the chiral extrapolation of lattice simulation results 
\cite{Leinweber:2005xz}
%\cite{Early,Moments,Young:2002cj,Young:2002ib,%
%Leinweber:2003dg,Young:2004tb,Leinweber:2004tc}, 
and establishing new approximation schemes to enable the extrapolation
of today's lattice QCD results \cite{Leinweber:2003dg,Young:2002ib}.

The established, model-independent approach to chiral effective field
theory is that of power counting, the foundation of chiral
perturbation theory (\chiPT).  However, this requires one to work in a
regime of pion mass where the next term in the truncated series
expansion makes a contribution that is negligible.  As there is no
attempt to model the higher-order terms of the chiral expansion, one
simply obtains the wrong answer if one works outside this region.
Knowledge of the power counting regime (PCR), where neglected
higher-order terms are truly small, is as important as the chiral
expansion itself.

Approximation schemes play a significant role in modern lattice QCD
simulations.  Consider for example the calculation of all-to-all
propagators \cite{Foley:2005ac}, where the most important low-lying eigenmodes
of the Dirac operator are treated exactly in the inversion process.
Having treated the dominant contributions precisely, the remainder of
the matrix inverse is approximated using stochastic estimator
techniques.  The FRR approach to \chiEFT\ also has these
characteristic features.  Just as the low-lying eigenmodes are treated
exactly, so is the chiral regime of the chiral expansion.  \FRRchiPT\
is mathematically equivalent to standard \chiPT\ to the finite order
one is working.\footnote{A survey of the literature reviewing
\chiEFT\ illustrates that most practitioners are unaware of this fact.
} Having treated the dominant contributions precisely, the remainder
of the chiral expansion is approximated using FRR-induced resummation
techniques.

Because terms of the chiral expansion beyond the finite order
calculated are treated in an approximate manner, \FRRchiPT\ is often
regarded as a model.  In the lattice community, models are usually
eschewed at all costs, but the costs are high.  Most chiral
extrapolations presented this year at Lattice '05, are still of the
most naive linear or polynomial form.  Those performing extrapolations
with traditional \chiPT\ are performing the extrapolations from well
outside the PCR.  The most common signature of this is that when
higher-order terms are calculated, they are almost always found to be
large, even in the favorable meson sector \cite{Bijnens:2005ne}.  If
one was working in the PCR to begin with, then the next order term of
the expansion is small by definition!  The reluctance to
quantitatively determine the PCR undermines the integrity and
credibility of lattice QCD predictions.

There continues to be a reluctant but growing recognition that some
form of resummation of the chiral expansion is necessary in order to
make contact with lattice simulation results of full QCD.  The
resummation of the chiral expansion induced through the introduction
of a finite-range cutoff in the momentum-integrals of meson-loop
diagrams is perhaps the best known resummation method
\cite{Leinweber:2005xz,Leinweber:2003dg,Young:2002ib}.
%--- for alternative proposals, see also
%Refs.~\cite{Djukanovic:2004px,Pascalutsa:2004je}.  
Taylor expansions of FRR fits to lattice QCD results for magnetic
moments indicate that terms to $m_\pi^{26}$ are required to reach the
first lattice data point at $m_\pi^2 = 0.2\ {\rm GeV}^2$
\cite{Young:2004tb}.  Given the astronomical number of low energy
constants to be determined if such calculations were even possible in
\chiPT, one must question if this really is the interesting physics.
% One is reminded of the case of
% thermodynamics, where knowing the momentum of each individual molecule
% as a function of time, reveals little of the relevant properties of
% temperature and pressure.

As we will demonstrate in the following, the quark masses accessible
with today's algorithms and supercomputers lie well outside the regime
of baryon \chiPT\ in its standard form.  This situation is unlikely to
change significantly until it becomes possible to directly simulate
QCD on the lattice within twice the squared physical mass of the pion
and with suitably large lattice volumes.  Still, one might wonder if
the lattice techniques that would allow simulations at light masses
within $2 m_\pi^2$, might also allow a calculation directly at the
physical pion mass, obliterating the chiral extrapolation problem
altogether.

\section{\FRRchiPT\ is not a model in the PCR}

To demonstrate that \FRRchiEFT\ is mathematically equivalent to
\chiPT\ to the order calculated and alleviate the myth that the
\FRRchiEFT\ approach is simply a model, we review the process of
renormalisation in a minimal subtraction scheme and in \FRRchiEFT.  To
leading one-loop order
\begin{equation}
M_N = a_0 + a_2 \mpi^2 + \chi_\pi I_\pi\, ,
\label{eq:MNlead}
\end{equation}
where $\chi_\pi = -{3\, g_A^2}/{(32\pi\fpi^2)}$ is the LNA coefficient
of the nucleon mass expansion, and $I_\pi$ denotes the relevant loop
integral. In the heavy baryon limit, this integral over pion momentum
is given by
\begin{equation}
I_\pi = \frac{2}{\pi} \int_0^\infty dk\, \frac{k^4}{k^2+\mpi^2} \, .
\label{eq:INpi}
\end{equation}
This integral suffers from a cubic divergence for large momentum. The
infrared behavior of this integral gives the leading nonanalytic
correction to the nucleon mass. This arises from the pole in the pion
propagator at complex momentum $k=i\mpi$ and will be determined
independent of how the ultraviolet behavior of the integral is
treated. Rearranging Eq.~(\ref{eq:INpi}) we see that the pole contribution can
be isolated from the divergent part
\begin{equation}
I_\pi = \frac{2}{\pi} \int_0^\infty dk\, \left(k^2-\mpi^2\right) 
    + \frac{2}{\pi} \int_0^\infty dk\, \frac{\mpi^4}{k^2+\mpi^2} \, .
\label{eq:IpiPole}
\end{equation}
The final term converges and provides $m_\pi^3$.
% %
% \begin{equation}
% \frac{2}{\pi} \int_0^\infty dk\, \frac{\mpi^4}{k^2+\mpi^2} = \mpi^3 \, ,
% \end{equation}
% %
% where we now recognize the choice of normalization of the loop
% integral, defined such that the coefficient of the LNA term is set to
% unity. 
% This choice is purely convention and allows for a much
% more transparent presentation of the differences in the chiral
% expansion with various regularization schemes.
%
In the most basic form of renormalization we could simply imagine
absorbing the infinite contributions arising from the first term in
Eq.~(\ref{eq:IpiPole}) into a redefinition of the coefficients $a_0$ and
$a_2$ in Eq.~(\ref{eq:MNlead}). This solution is simply a minimal subtraction
scheme and the renormalized expansion can be given without making
reference to an explicit scale,
\begin{equation}
M_N = c_0 + c_2 \mpi^2 + \chi_\pi \mpi^3 \, ,
\label{eq:MSrenLNA}
\end{equation}
with the renormalized coefficients defined by
\begin{equation}
c_0 = a_0 + \chi_\pi \frac{2}{\pi} \int_0^\infty dk\, k^2 \, ,\quad
c_2 = a_2 - \chi_\pi \frac{2}{\pi} \int_0^\infty dk\, .
\end{equation}
Equation~(\ref{eq:MSrenLNA}) therefore encodes the complete quark mass
expansion of the nucleon mass to ${\cal O}(\mpi^3)$. 
This result will be precisely equivalent to any form of minimal
subtraction scheme, where all the ultraviolet behavior is absorbed
into the two leading coefficients of the expansion.
Such a minimal subtraction scheme is characteristic of the commonly
implemented dimensional regularization.

We now describe the chiral expansion within finite-range
regularization, where the cut-off scale remains explicit. In
particular, we highlight the mathematical equivalence of FRR and
dimensional regularization in the low energy regime. We introduce a
functional cutoff, $u(k)$, defined such that the loop integral is
ultraviolet finite,
\begin{equation}
I_\pi = \frac{2}{\pi} \int_0^\infty dk\, \frac{k^4\, u^2(k)}{k^2+\mpi^2} \, .
\label{eq:INpiFRR}
\end{equation}
To preserve the infrared behavior of the loop integral, the regulator
is defined to be unity as $k\to 0$.  For demonstrative purposes, we
choose a dipole regulator $u(k)=(1+k^2/\Lambda^2)^{-2}$, giving
\begin{equation}
I_\pi^{\rm DIP} =
\frac{\Lambda^5(\mpi^2+4\mpi\Lambda+\Lambda^2)}{16(\mpi+\Lambda)^4} 
\,\sim \frac{\Lambda^3}{16} - \frac{5\Lambda}{16}\mpi^2 
+ \mpi^3 - \frac{35}{16\Lambda} \mpi^4 + \ldots \,,
\label{eq:INpiDIP}
\end{equation}
The first few terms of the Taylor series expansion, as shown, provide
the relevant renormalisation of the low-energy terms.
The renormalized expansion in FRR is therefore precisely equivalent to
Eq.~(\ref{eq:MSrenLNA}) up to ${\cal O}(\mpi^3)$ where the leading
renormalized coefficients are given by
\begin{equation}
c_0 = a_0 + \chi_\pi \frac{\Lambda^3}{16} \, ,\quad
c_2 = a_2 - \chi_\pi \frac{5\Lambda}{16}  \, .
\label{LambdaFree}
\end{equation}
As $a_0$ and $a_2$ are fit parameters, the value $\Lambda$ takes is
irrelevant and plays no role in the expansion to the order one is
working; in this case $m_\pi^3$.  
Hence the suggestion, for example, that infrared regularization is
somehow less model dependent than FRR is false and misleading.
%
% The original ideas leading to finite-range cut-off schemes emphasized
% the importance of suppressing large-momenta in the loop integrals
% involving low-energy effective degress of freedom, as the
% contributions would be incorrect.  No matter how compelling these
% arguments might be, they have absolutely nothing to do with \chiPT.
Within the PCR of \chiPT\ there is no physics in the regulator.

% Systematic improvement of the FRR expansion is also possible.  
%
It is straight forward to extend this procedure to next-to-leading
nonanalytic order, explicitly including all terms up to $m_q^2\sim
\mpi^4$.  Most importantly, there are nonanalytic contributions of
order $\mpi^4\log\mpi$ arising from the $\Delta$-baryon and tadpole
loop contributions.  Details may be found in
\cite{Leinweber:2005xz,Leinweber:2003dg}.

\section{Power-counting regime (PCR)}

The PCR is the regime in which neglected higher-order terms of the
standard expansion of \chiPT\ are small, because $m_\pi$ is a small
number raised to a high power.  Since the chiral expansion of
\chiPT\ is truncated with no attempt to estimate the contribution of
higher-order terms, 
% knowing the PCR of a given truncation is
% absolutely essential to extracting correct physics from \chiPT.  One
one
simply obtains the wrong answer if one works outside the PCR.

As discussed in detail surrounding Eq.~(\ref{LambdaFree}), the FRR
chiral expansion is mathematically equivalent to that of \chiPT\ to
the finite order one is working.  In other words, these terms of the
FRR expansion are independent of the regulator parameter, $\Lambda$.
Thus \FRRchiEFT\ can be used to determine the power-counting regime by
varying $\Lambda$ and identifying the regime in pion mass
in which the results are invariant to some level of precision.

\begin{figure}[!t]
\begin{center}
\includegraphics[width=7.5cm]{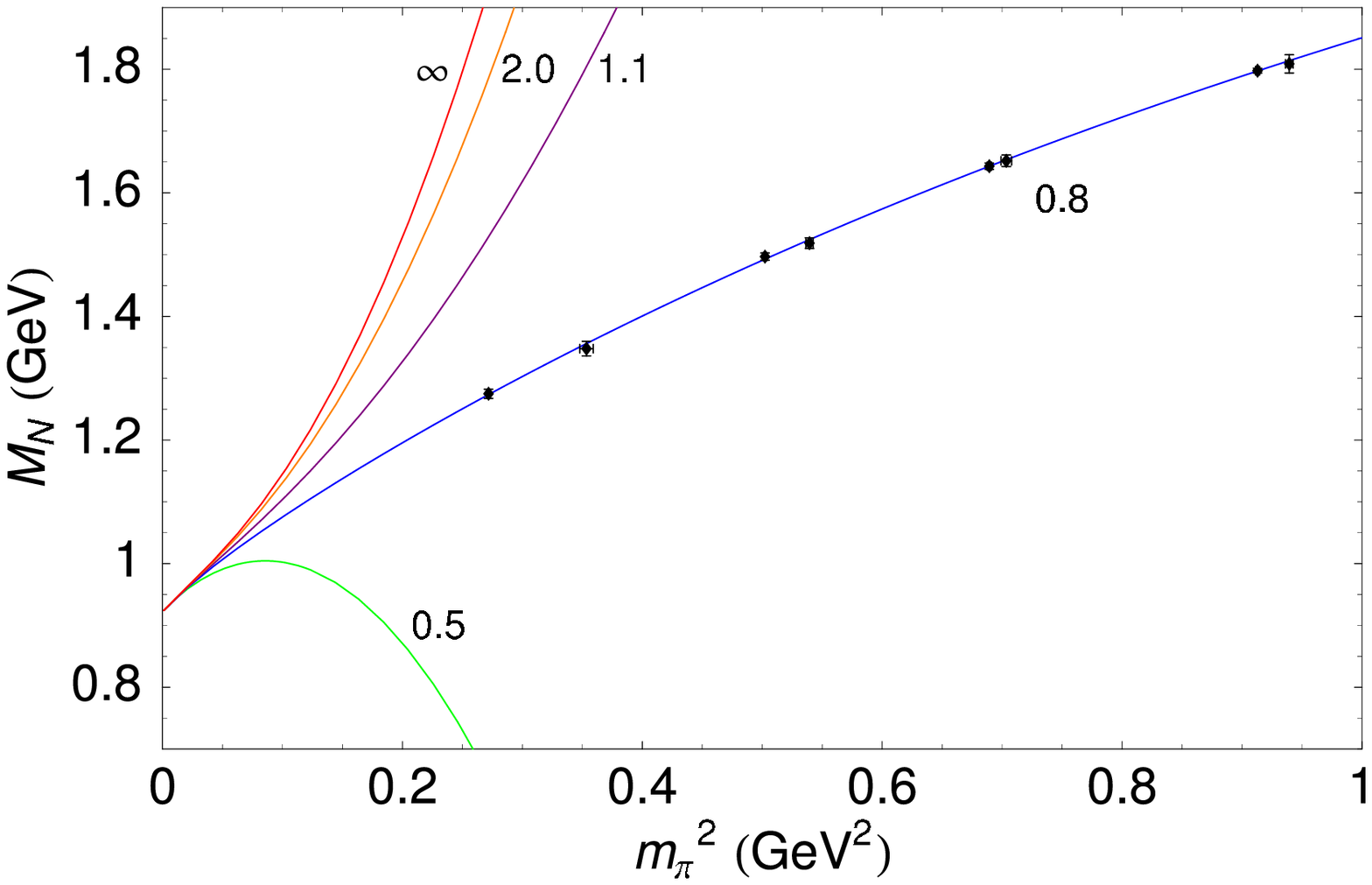}%
\includegraphics[width=7.5cm]{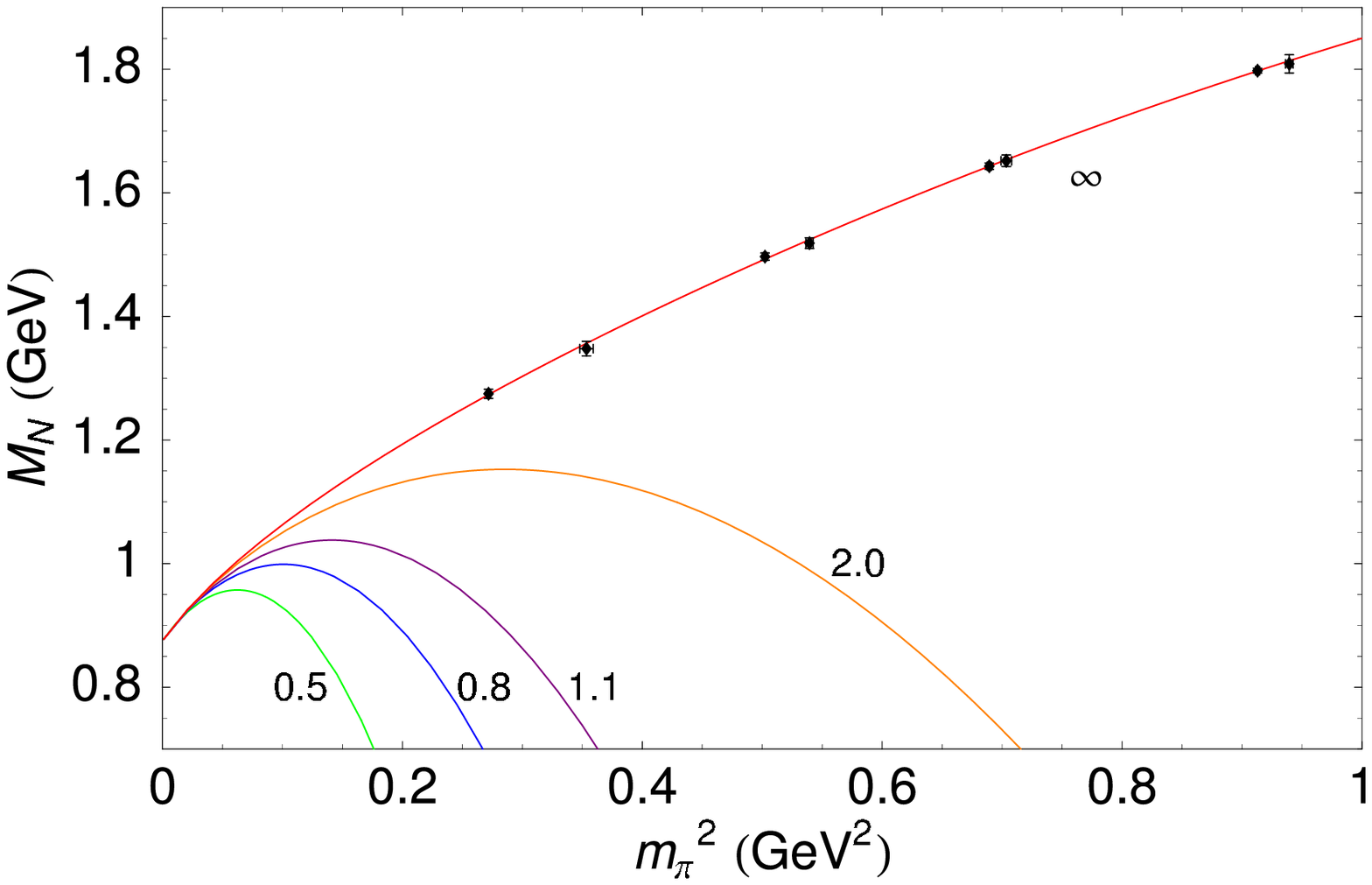}
\end{center}
\vspace{-0.7cm}
\caption{With the low-energy parameters $c_0$, $c_2$ and $c_4$ fixed
to those obtained by the fit to lattice data with $\Lambda = 0.8$
(left) and the minimal subtraction limit $\Lambda \to \infty$ (right),
the chiral expansion is shown for various values of the dipole
regulator scale, $\Lambda=0.5,\ 0.8,\ 1.1,\ 2.0\ {\rm and}\
\infty\ \gev$.
\vspace{-0.3cm}
}
\label{fig:fixC}
\end{figure}

\begin{figure}[t]
\begin{center}
\includegraphics[width=7.5cm]{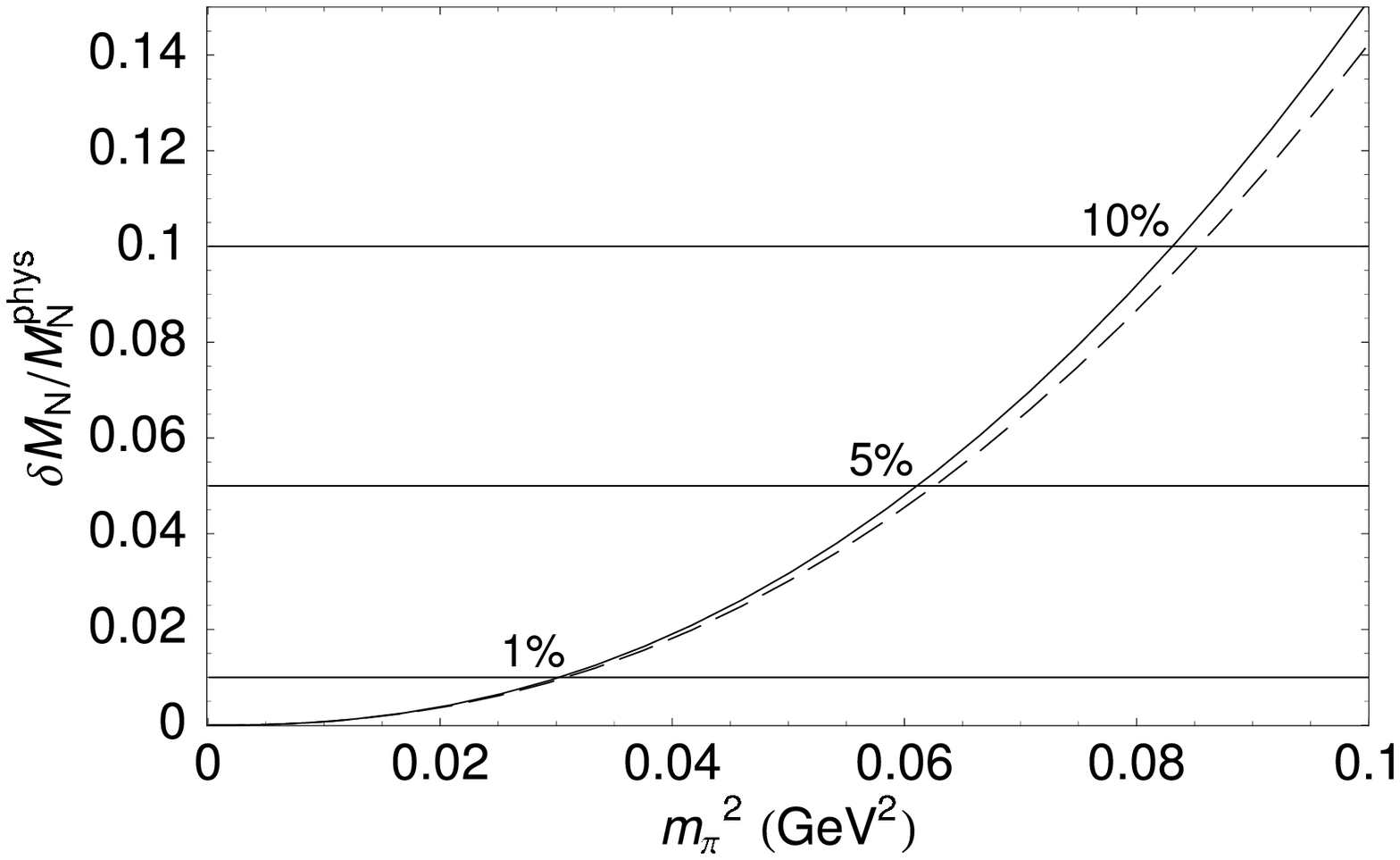}%
\includegraphics[width=7.5cm]{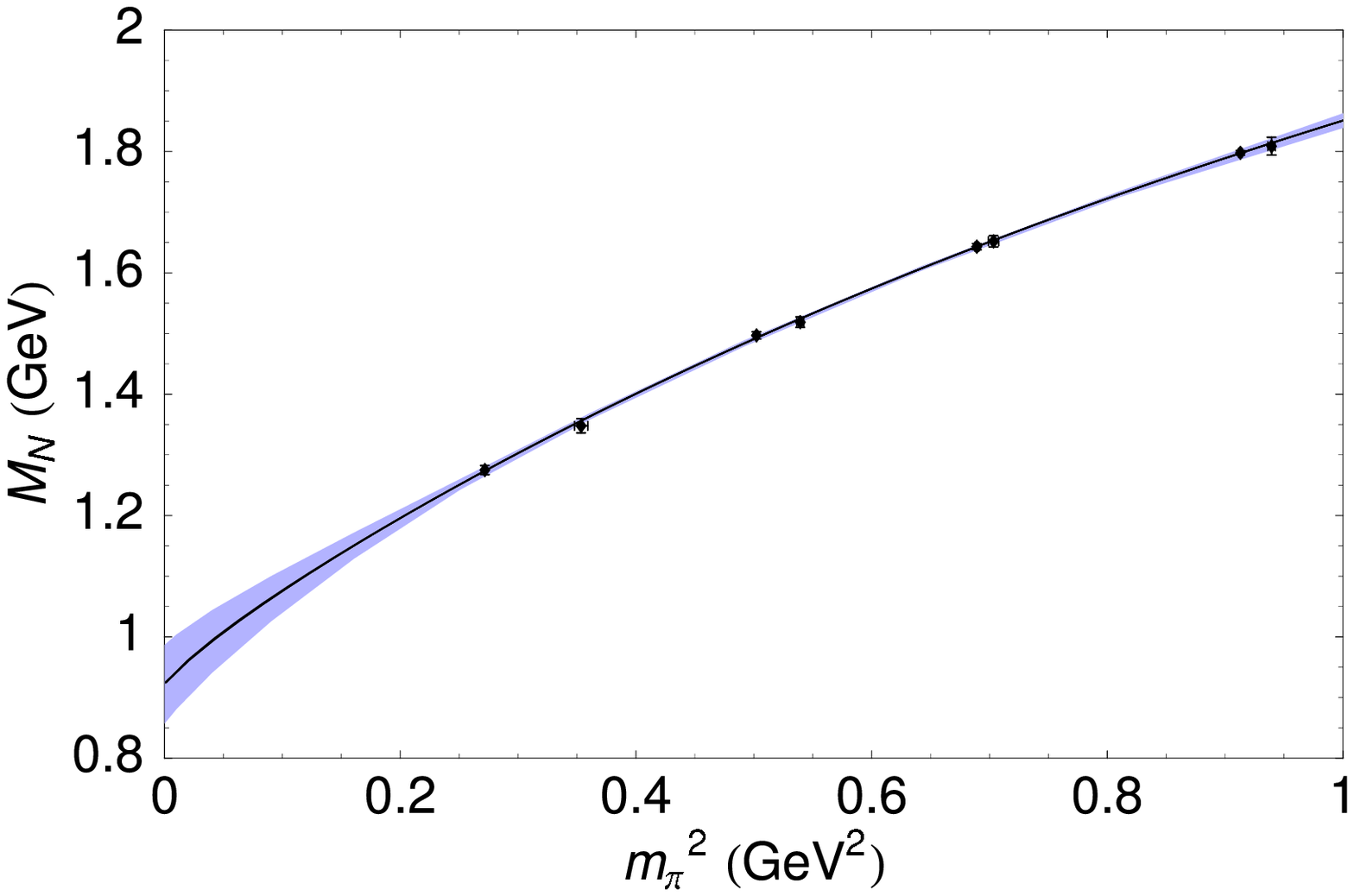}
\end{center}
\vspace{-0.7cm}
\caption{(left) For fixed low-energy coefficients $c_0$, $c_2$ and
  $c_4$, the relative difference in the nucleon mass expansion for two
  extremal regularization scales, where $\delta
  M_N=M_N(\Lambda=\infty)-M_N(\Lambda=0.5)$.  The solid and dashed
  curves correspond to the differences displayed in the left and
  right-hand panels of Fig.~\protect\ref{fig:fixC} where the $c_i$ are
  determined with a dipole of scale $\Lambda=0.8$ and $\infty\ \gev$,
  respectively.
(right) Extrapolation of CP-PACS collaboration simulation results
  \protect\cite{AliKhan:2001tx} to the chiral limit using finite-range
  regularization \cite{Leinweber:2003dg}.  Differences between the
  illustrated dipole, monopole, Gaussian and theta-function regulators
  cannot be resolved on this scale.  The one-standard deviation error
  bound for the dipole extrapolation is also illustrated.
}
\label{fig:pcregime}
\end{figure}

Fig.~\ref{fig:fixC} illustrates the fourth-order chiral expansion for
various dipole regulator parameters $\Lambda$.
%, ranging between $0.5\gev$ and $\infty$.  
Since the expansion
% renormalized coefficients and nonanalytic terms 
to fourth order is automatically independent of $\Lambda$, the
observed changes in the curves are simply a reflection of the changes
in terms beyond fourth order.
%
% The power counting regime can be identified as the regime where the
% curves are independent of $\Lambda$ at some level of precision.
Fig.~\ref{fig:pcregime} displays the relative error between the two
extremal regularization scales for the left (solid) and right (dashed)
panels of Fig.~\ref{fig:fixC}.  The regime where the curves
agree within one percent is $m_\pi \le 180$ MeV extending only 40 MeV
beyond the physical mass.  While this is excellent news for
understanding experimental results within chiral perturbation theory,
it also illustrates that today's naive application of \chiPT\ to the
chiral extrapolation problem in lattice QCD is inappropriate.

%@@@@@@@@@@@@@@@@@@@@@@@@@@@@@@@@@@@@@@@@@@

\section{\FRRchiEFT\ as a solution to the chiral
   extrapolation problem.}

To investigate the extent to which various regulators provide a
model-independent estimator for the {\it sum} of higher-order terms of
the chiral expansion, beyond the finite order calculated, the
finite-range regulator $u(k)$ is taken to be either a sharp
theta-function cut-off, a dipole, a monopole or finally a Gaussian.
These regulators have very different shapes, with the only common
feature being that they suppress the integrand for momenta greater
than $\Lambda$.
Figure~\ref{fig:pcregime} (right) displays the extrapolation
\cite{Leinweber:2003dg} of full QCD simulation results of the nucleon
mass from the CP-PACS collaboration \cite{AliKhan:2001tx} using
\FRRchiEFT\ to fourth order in the expansion; {\it i.e.}  to order
$m_\pi^4 \log m_\pi$.  The curves are indistinguishable and produce
physical nucleon masses which differ by less than 0.1\%.

The astonishing discovery in FRR chiral effective field theory, is
that the term-by-term details of the higher-order chiral expansion are
largely irrelevant in describing the chiral extrapolation of
simulation results.  The coefficients of the higher-order terms
($m_\pi^5$ and beyond) appearing in the FRR expressions differ
significantly, yet the curves of Fig.~\ref{fig:pcregime} (right) are
indistinguishable.  Given the level of agreement between the curves
associated with different regulators, and the fact that the lattice
results are described perfectly, it is sufficient to approximate the
remainder of the chiral expansion in terms of a single parameter,
$\Lambda$.

\section{Summary}

So why does \FRRchiEFT\ work?  The essential physics is that loop
integrals vanish as the quark masses grow large.  Exactly how zero is
approached is governed by the regulator parameter, $\Lambda$, and in
most cases $\Lambda$ is constrained by lattice QCD simulation results.
The contribution of any individual higher-order term is largely
irrelevant.  The only thing that really counts is that there are other
terms that enter to ensure the sum of all terms of the loop integral
approaches zero, in accord with what is observed in lattice QCD
calculations.  Of course, this beautiful feature of FRR expansions
would be lost if one were to truncate the expansion at any finite
order.  Resummation of chiral effective field theory is essential to
solving the chiral extrapolation problem.

The finite-range regularisation (FRR) approach to chiral effective
field theory ($\chi$EFT) provides an approximation scheme that
connects {\it today's} lattice simulation results to the physical
world.  It has been successfully applied to describe
partially-quenched simulation results of the rho meson mass
% using partially-quenched \FRRchiEFT, 
in a unified analysis incorporating both finite volume and finite
lattice spacing artifacts \cite{Allton:2005fb}.  The CSSM lattice
collaboration has completed extensive simulations of baryon
electromagnetic form factors.  An associated quenched \FRRchiEFT\
analysis of the magnetic moments correcting finite-volume and quenched
artifacts has led to the most precise determination of the nucleon's
strange magnetic moment \cite{Leinweber:2004tc}.

\end{document}